\documentclass[twocolumn,aps,showpacs,floatfix,prc]{revtex4}
\usepackage[dvips]{epsfig}
\newcommand{\snn}{\sqrt{s}_{NN}}
\newcommand{\la}{\langle}
\newcommand{\ra}{\rangle}
\begin{document}

\title{$J/\psi$ suppression and QCD phase diagram }
 
\author{A. K. Chaudhuri}
\email[E-mail:]{akc@veccal.ernet.in}
\author{Partha Pratim Bhaduri}
\email[E-mail:]{partha.bhaduri@veccal.ernet.in}
\affiliation{Variable Energy Cyclotron Centre,\\ 1/AF, Bidhan Nagar,
Kolkata 700~064, India}

\begin{abstract}

QCD phase diagram is obtained by analysing centrality dependence of $J/\psi$ suppression in $\snn$=17.3 Pb+Pb and $\snn$=200 GeV Au+Au collisions. $J/\psi$'s produced in initial interactions are assumed to dissolve if local temperature exceeds a threshold temperature. The threshold temperature depends on   the (local) fluid temperature and baryonic chemical potential, which are obtained, under certain assumptions,  from experimentally determined quantities e.g. rapidity density, net baryon density. 
QCD critical line with  curvature parameter $\kappa\approx 0.018-0.23$ is consistent with experimental $J/\psi$ data. $\kappa$ is factor of $\sim$3 larger than in   lattice QCD calculations. 

  \end{abstract}  
\maketitle
 
\section{introduction} 

In recent years, there is much interest in QGP phase diagram in $T-\mu$ plane, $T$ the temperature and $\mu$ the baryonic chemical potential  \cite{Fukushima:2010bq},\cite{Fukushima:2008wg},\cite{Stephanov:2007fk},\cite{Asakawa:1989bq}. Presently, only one point in the phase diagram is known from lattice QCD simulations, at $\mu \approx$0, the QCD transition is a cross over \cite{Aoki:2006we} at (pseudo) critical temperature $T=T_c\approx$157 (3)(3) MeV \cite{Borsanyi:2011zz}.
From theoretical considerations, QCD phase transition is expected to be 1st order in baryon dense matter.  One then expects a QCD critical end point (CEP), where the transition changes from 1st order to cross-over. 
 To access the QCD phase diagram and CEP, 
systematic study of nuclear collisions, as a function of collision energy, has been planned at   FAIR \cite{Senger:2011zz} in the energy range $E_{lab}$=5-40 GeV.

\begin{figure}[t]
\center
 \resizebox{0.35\textwidth}{!}{%
  \includegraphics{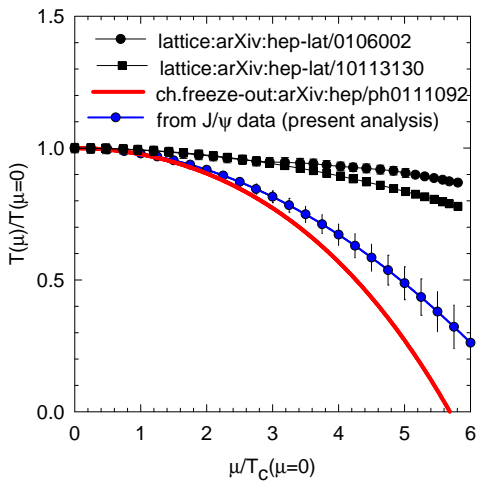}
}
\caption{(color online)Lattice QCD calculation of QCD phase diagram, in  imaginary chemical potential method \cite{Fodor:2001pe}  and Taylor expansion method \cite{Kaczmarek:2011zz}  are shown. The red line is the chemical freeze-out curve obtained in a statistical model \cite{Cleymans:2006qe}. The   phase diagram obtained in the present analysis is shown as the blue circles.}
  \label{F1}
\end{figure}

At finite baryon density, Fermion determinant is complex and standard technique of Monte-Carlo importance sampling fails. Several techniques have been suggested to circumvent the problem, (i) reweighting \cite{Fodor:2001pe},\cite{Fodor:2004nz} , (ii) analytical continuation of imaginary chemical potential \cite{de Forcrand:2002ci}, \cite{D'Elia:2002gd} and (iii)Taylor expansion \cite{Allton:2003vx},\cite{Gavai:2003mf}. These methods has been used to locate the phase boundary in $T-\mu$ plane. The calculations suggest that   the curvature parameter ($\kappa$)
 in the expansion,

\begin{equation} \label{eq1}
\frac{T_c(\mu)}{T_c(\mu=0)}=1-\kappa \left (\frac{\mu}{T_c(\mu=0)} \right )^2
\end{equation}

\noindent is small    \cite{Philipsen:2008gf}. As an example, in Fig.\ref{F1},
QCD phase diagram obtained in the analytical continuation method \cite{Fodor:2001pe} (the filled circles) and in Taylor expansion    \cite{Kaczmarek:2011zz} of (chiral) are shown. Both the methods gives nearly identical phase diagram for $\mu/T_c(\mu=0) < 3 GeV$, curvature parameter is small, $\kappa\approx 0.006$. At larger $\mu$, they differ marginally. For comparison, in Fig.\ref{F1}, chemical freeze-out curve \cite{Becattini:2005xt},\cite{Cleymans:2006qe}, obtained in statistical model analysis of particle ratios are shown (the red line). Curvature of the chemical freeze-out curve is factor of 4 larger than the curvature in the QCD phase diagram. Small curvature of the QCD critical line, compared to the chemical freeze-out is interesting. Experimental signal of critical end point will get diluted as the deconfined medium produced at the critical end point will evolve longer to reach chemical freeze-out. Fluid will have more time to washout any signature of CEP.

In the present letter, we obtain the curvature parameter $\kappa$ by analysing experimental data on $J/\psi$ suppression.  Existing data in $\sqrt{s}_{NN}$=17.3 GeV Pb+Pb and $\sqrt{s}_{NN}$=200 GeV Au+Au collisions, constrain the curvature parameter within a narrow range, $\kappa\approx 0.018-0.023$. It is factor of $\sim$ 3 larger than in the lattice QCD calculations. To our knowledge, this is the first attempt to obtain QCD critical line from experiment data.

 \section{Model for $J/\psi$ suppression in QGP}
 
 In a deconfined medium (QGP), inter-quark (color) potential is screened. The screening increases with temperature and above a certain temperature, the potential is too weak to bind charm and anti-charm quarks into a bound state. Charmonium states are then suppressed. 
Matsui and Satz  \cite{Matsui:1986dk}, first predicted the phenomena.
  Over  the  years, in several experiments,   $J/\psi$ yield in heavy ion collisions has been measured. Measurements at SPS ($\snn$=17.3 GeV) and RHIC ($\snn$=200 GeV) energy validate Matsui and Satz's predictions. 
  

 To mimic the onset of deconfining phase transition above a critical energy density and subsequent melting of $J/\psi$'s, QGP motivated  threshold model was proposed in  \cite{Blaizot:2000ev,Blaizot:1996nq}. In the threshold model, $J/\psi$ suppression is linked with the local transverse density. If the local transverse density, at the point where $J/\psi$ is formed, exceed a critical or threshold value, $J/\psi$'s are melted. In the present paper, we use a variant of the threshold model,
 if the local temperature of the fluid, exceed a threshold value $T_{J/\psi}$,   $J/\psi$'s are melted.  
Approximately  30-40\% of observed  $J/\psi$'s are  from feed-down decay of the higher states, $\psi^\prime$ and $\chi$ \cite{Satz:2005hx}. Higher states, $\psi^\prime$ and $\chi$ are expected to melt at temperature $T_{\psi^\prime}$,$T_\chi$ $< T_{J/\psi}$. 
 Lattice QCD calculations indicate that    $J/\psi$'s can survive up to $T_{J/\psi}=1.5-2T_c$ \cite{Satz:2005hx}, higher states survive up to $T_{\psi^\prime}\approx T_\chi\approx1-1.2T_c$. For sequential melting of charmonium states, if the melting temperatures $T_{J/\psi}$ and $T_{\psi^\prime},T_{\chi} $ are substantially different, $J/\psi$ survival probability should show at least two step distribution (see for example Fig.4 in  \cite{Digal:2001bh}). No such two step distribution is seen either in $\snn$=200 GeV Au+Au or in $\snn$=17.3 GeV Pb+Pb collisions. Absence of two step structure in $J/\psi$ survival probability indicate that the melting temperatures of $J/\psi$'s and higher states are not substantially different.
 In the following we assume direct and feed-down $J/\psi$'s are melted at the common temperature $T_{J/\psi}=KT_c$. In a baryon rich medium, melting temperatures $T_{J/\psi}$  depend on the QGP phase diagram through the critical temperature $T_c(\mu)$ as in Eq.\ref{eq1}.

Threshold model require the local temperature ($T_i$) and (baryonic) chemical potential ($\mu_i$) of the fluid produced in nuclear collisions. We propose to obtain them from experimentally measured quantities, e.g. charged particles rapidity density $\frac{dN_{ch}}{d\eta}$ and net baryon rapidity density $(B-\bar{B})\frac{dn}{dy}$.  Experimentally measured values of $\frac{dN_{ch}}{d\eta}|_{\eta=0}$ \cite{Adler:2004zn} and $(B-\bar{B})\frac{dn}{dy}|_{y=0}$ \cite{Appelshauser:1998yb},\cite{Bearden:2003hx}, in 0-5\% 
$\snn$=17.3 and 200 GeV collisions,  are listed in table.\ref{table1}. The initial entropy density is then determined from the well known relation \cite{Hwa:1985xg},

\begin{equation} \label{eq2}
 \la s \ra _i=\frac{1.5\times 3.6}{\pi R^2 \tau_i}   \frac{dN_{ch}}{d\eta} \vert_{\eta=0}
 \end{equation}

In Eq.\ref{eq2}, $\tau_i$ is the initial time, for which we have used the canonical value $\tau_i$=1.0 fm.  We do note that Eq.\ref{eq2} is obtained under the assumption of isentropic, longitudinal expansion. Similarly, initial baryon density is   obtained  as,

\begin{equation}\label{eq3}
 \la n \ra _i=\frac{1}{\pi R_A^2\tau_i}(B-\bar{B})\frac{dn}{dy}\vert_{y=0}
\end{equation}

$\la .. \ra$ in Eq.\ref{eq2}-\ref{eq3} indicate that the entropy density   and baryon density in Eqs.\ref{eq2},\ref{eq3}  must be understood as spatially averaged over entropy/baryon density.
To introduce spatial dependence, we assume, that
in an impact parameter $b$ collision, initial  entropy density and baryon density in the transverse plane is distributed as \cite{Kolb:2003dz},

 \begin{table}[t] 
\caption{\label{table1} Rapidity density $\frac{dN_{ch}}{d\eta}$, net baryon density $(B-\bar{B})\frac{dn}{dy}$ in 0-5\% $\snn$=17.3 Pb+Pb and $\snn$=200 GeV Au+Au collisions. Corresponding average temperature ($T_i$) and  $\mu_i$ are also listed. 
} 
\begin{ruledtabular} 
  \begin{tabular}{|c|c|c|c|c|}\hline
$\snn$ & $\frac{dN_{ch}}{d\eta}\vert_{\eta=0}$  & $(B-\bar{B})\frac{dn}{dy}\vert_{y=0}$  & $\la T\ra_i$ & $\la \mu\ra_i$ \\   
(GeV)  &   &    & (MeV) & (MeV) \\  \hline
 17.3 (Pb+Pb)   &308   & 75   & 191.5 & 390.2 \\  \hline
200.0 (Au+Au)  &690   & 7   & 242.4 & 23.6 
 \end{tabular}\end{ruledtabular}  
\end{table}

\begin{eqnarray} 
s_i(x,y) &=& s_0\left [ (1-f) \frac{N_{part}(x,y)}{2}+ f N_{coll}(x,y)\right ],  \label{eq4} \\
n_i(x,y) &=& n_0 \left [ (1-f) \frac{N_{part}(x,y)}{2}+ f N_{coll}(x,y)\right ] \label{eq5}
\end{eqnarray}

\noindent
 $s_0$ and $n_0$ is fixed to reproduce the average entropy density $\la s \ra _i$ and baryon density $\la n\ra_i$, as determined from experimental values.  
 $N_{part}(x,y)$ and $N_{coll}(x,y)$ in Eq.\ref{eq4},\ref{eq5}  are the transverse profile of the participant density and binary collision number, obtained in a Glauber model.  $f$ is the fraction of hard scattering.
 It is approximately constant   in the energy range $\snn$=17-200 GeV, $f\approx$0.12.

Local entropy density ($s_i(x,y)$) and baryon density ($n_i(x,y)$) can be converted into local temperature and baryonic chemical potential using an equation of state.
In  \cite{Kapusta:2010ke}, a parametric form of QGP equation of state was given, which matches with the lattice QCD simulation at zero chemical potential   and to the known properties of nuclear matter at zero temperature.   Entropy density ($s$) and  baryon density ($n$)   in parameterised form are,

\begin{eqnarray}  
s&=&\frac{4\pi^2}{90}\left (16+\frac{21N_f}{2}\right )T^3+\frac{N_f}{9}\mu^2T  -  2CT   \label{eq7}\\
n&=& \frac{N_f}{9}\mu T^2+\frac{N_f}{81\pi^2}\mu^3  - 2D\mu^2 \label{eq8}
\end{eqnarray}

$N_f\approx 2.5$, $2C\approx 0.24$, $D\approx0$.
In table.\ref{table1}, the average initial temperature and baryonic chemical potential   in $\snn$=17.3GeV Pb+Pb and $\snn$=200 GeV Au+Au collisions are listed. From SPS to RHIC, chemical potential decreases by a factor of $\sim$15. Temperature however is increased by  $\sim$25\%.

To obtain $J/\psi$ survival probability ($S_{J/\psi}$), we randomly distribute $J/\psi$'s   in the reaction plane and   count the number of  $J/\psi$'s that survive the threshold condition, $T_{local} < T_{J/\psi}$. 
We note that apart from suppression in deconfined matter, $J/\psi$'s are suppressed in cold nuclear matter also (CNM effect).  It is important to eliminate the suppression due to CNM effect. If cold nuclear matter effects are accounted for, the survival probability of $J/\psi$'s can be obtained as,

\begin{equation}\label{eq10}
S_{J/\psi}=S_{QGP}\times S_{CNM}
\end{equation}

  \begin{figure}[t]
\center
 \resizebox{0.35\textwidth}{!}{%
  \includegraphics{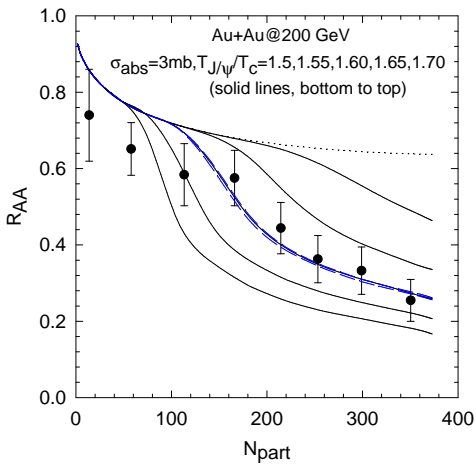}
}
\caption{(color online)The filled circles are  PHENIX data \cite{Adare:2006ns} for nuclear modification factor in  $\sqrt{s_{NN}}$=200 GeV Au+Au collisions. Black lines   (from bottom to top) are $j/\psi$ survival probability in the present model   for $\kappa$=0.03 and $J/\psi$ melting   temperature $T_{J/\psi}=KT_c$, K=1.5, 1.55, 1.6, 1.65 and 1.7, $T_c$=0.16.  The dashed blue lines are prediction obtained with fixed $T_{J/\psi}=1.65T_c$ and   $\kappa$=0 and 0.1 and. The dotted line is the suppression due to the CNM effect.}
  \label{F2}
\end{figure}  

\noindent where $S_{CNM}$ is the survival probability in cold nuclear matter and $S_{QGP}$ is the survival probability in QGP. In $\sqrt{s}_{NN}$=200 GeV d+Au collisions,  $J/\psi$ production is consistent with cold nuclear matter effect quantified in a Glauber model of nuclear absorption with $\sigma_{abs}=2\pm 1$ mb \cite{Vogt:2005ia}. $\sigma_{abs}\approx$ $7.6\pm 0.07$ mb  \cite{Arnaldi:2009ph} in $\snn$=17.3 GeV Pb+Pb collisions.

\section{Results}

The present model has two parameter, the threshold temperature for $J/\psi$ melting, 
$T_{J/\psi}=K T_c$ and the curvature parameter  $\kappa$ of the QCD critical line. The critical temperature is fixed to  $T_c(\mu=0)$=160 MeV.
The free parameters are obtained by fitting   experimental data on $J/\psi$ production in $\snn$=17.3 GeV Pb+Pb \cite{Arnaldi:2009ph} and $\snn$=200 GeV Au+Au \cite{Adare:2006ns} collisions. 
In $\snn$=200 GeV Au+Au collisions, the fluid is essentially baryon free and
model predictions do not depend sensitively on the curvature parameter. 
The dependence on melting temperature however is strong. In Fig.\ref{F2}, for threshold temperature $T_{J/\psi}/T_c$=1.5, 1.55,1.6, 1.65, and 1.7, model predictions are compared with the PHENIX data \cite{Adare:2006ns} for nuclear modification factor in $\snn$=200 GeV Au+Au collisions. 
CNM effect is quantified in Glauber model of nuclear absorption with $\sigma_{abs}$=3 mb. Experimental data   are well explained for the threshold temperature  $\frac{T_{J/\psi}}{T_c}=  1.60\pm 0.03$.  The melting temperature $T_{J/\psi}\approx 1.60  T_c$ is consistent with lattice QCD calculation  \cite{Satz:2005hx}.  The blue lines in Fig.\ref{F2} are model predictions with $\kappa$=0.0 and 0.1, $T_{J/\psi}=1.6T_c$. They cannot be distinguished.
  
\begin{figure}[t]
\center
 \resizebox{0.35\textwidth}{!}{%
  \includegraphics{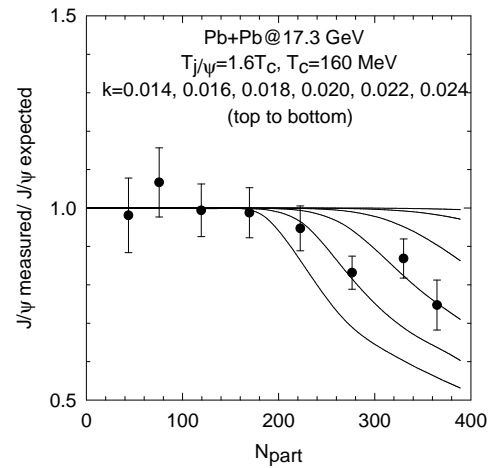}
}
\caption{ The solid circles are NA60 data for measured $J/\psi$ over expected $J/\psi$ in 158 AGeV Pb+Pb collisions. The solid lines (top to bottom) are present model predictions with $\kappa$=0.014, 0.016, 0.018, 0.020, 0.022 and 0.024 
respectively. $T_{J/\psi}=1.60T_c$, $T_c$=160 MeV.}
  \label{F3}
\end{figure}

Curvature parameter is constrained when $J/\psi$'s are absorbed in baryon rich plasma. In 158 AGeV Pb+Pb collisions fluid is baryon rich and $J/\psi$ suppression depend sensitively on the curvature parameter $\kappa$.
In Fig.\ref{F3}, centrality dependence of $J/\psi$ suppression in $\snn$=17.3 GeV Pb+Pb collisions is shown  \cite{Arnaldi:2009ph},\cite{Brambilla:2010cs}. The data are shown as the ratio of measured $J/\psi$'s over the expected $J/\psi$'s (in a Glauber model of nuclear absorption).
Upto $N_{part}$=200, the ratio is consistent with unity. Suppression in addition to the CNM effect  is required in $N_{part}>200$ collisions only. 
In Fig.\ref{F3}, the solid lines are threshold model predictions   with $T_{J/\psi}=1.60T_c$ and $\kappa$=0.014, 0.016, 0.018, 0.020, 0.022 and 0.024 respectively.   Unlike in $\snn$=200 GeV Au+Au collisions, in $\snn$=17.3 GeV Pb+Pb collisions, $J/\psi$ survival probability shows sensitive dependence on $\kappa$.  
For $\kappa\leq 0.016$, $J/\psi$'s are not suppressed in QGP medium.  
Best description to the data is obtained with   $\kappa \approx$0.02 ($\chi^2/N\approx 1.0$). Description to the data deteriorate for lower or higher curvature. If uncertainty in the melting temperature   is   accounted for, Pb+Pb data fix the
curvature parameter within a narrow range,  $\kappa= 0.018-0.023$,  for $T_{J/\psi}=(1.57-1.63)T_c$.

In Fig.\ref{F1},  QCD critical line obtained with the curvature parameter $\kappa= 0.018-0.023$ is shown (the blue circles with error bars). 
Curvature $\kappa=0.018-0.023$, obtained from $J/\psi$ data   is  $\sim$ 3 larger than in lattice QCD simulations $(\kappa\approx 0.006$). It is also  closer  to the chemical freeze-out curve.  The result is encouraging with respect to detection of QCD critical end point. If QCD critical line is close to the chemical freeze-out,  possibility of distortion of signals associated with CEP, e.g. charge fluctuations etc.,  will be lessened. 

Before we summarise the results some limitations of the model are noted.    Experimental information of particle density and baryon density is converted to local temperature and baryonic chemical potential using the EOS (see Eq.\ref{eq7},\ref{eq8}). The   EOS is for massless quarks and gluons,   adjusted to reproduce zero chemical potential lattice QCD simulations. How far it represent the real world is uncertain. The model is also a static model.
It neglect the possibility that $J/\psi$'s, initially produced  in a cold region
and survived, at a later time
can drift into a hotter region  and get melted. Such possibilities, if accounted for would reduce the survival probability. Consequently,  the present model give a lower limit of curvature parameter $\kappa$. 
 
\section{summary and conclusions}

To summarise, we have shown that in a threshold model,  
$J/\psi$ suppression is sensitive to the curvature parameter of the QCD phase diagram. In threshold model, $J/\psi$ suppression depend on the local temperature and chemical potential. A model is proposed to obtain local temperature and chemical potential, from experimentally determined quantities e.g. rapidity density, net baryon rapidity density. QCD critical line with curvature parameter $\kappa$=0.018-0.023 is consistent with  experimental centrality dependence of $J/\psi$ suppression in $\snn$=17.3 GeV Pb+Pb and $\snn$=200 GeV Au+Au collisions. $\kappa$=0.018-0.023 is approximately 3 times larger than that obtained in lattice QCD simulations. 
  
 
\end{document}